\documentclass[lettersize,journal]{IEEEtran}
\usepackage{amsmath,amsfonts}
\usepackage{algorithmic}
\usepackage{algorithm}
\usepackage{array}
\usepackage[caption=false,font=normalsize,labelfont=sf,textfont=sf]{subfig}
\usepackage{textcomp}
\usepackage{stfloats}
\usepackage{url}
\usepackage{verbatim}
\usepackage{graphicx}
\usepackage{cite}
\usepackage{xcolor}
\usepackage{multirow}
\usepackage{subcaption} 
\usepackage{enumitem}
\usepackage{threeparttable}
\usepackage{booktabs}
\usepackage{balance}
\definecolor{myblue}{rgb}{0.2, 0.4, 0.8}
\begin{document}

\title{EGRA: Toward Enhanced Behavior Graphs and Representation Alignment for Multimodal Recommendation}

\author{Xiaoxiong Zhang, Xin Zhou, Zhiwei Zeng, Yongjie Wang, Zhiqi Shen ~\IEEEmembership{Senior Member,~IEEE}

\thanks{Xiaoxiong Zhang, Xin Zhou, Zhiwei Zeng, Yongjie Wang and Zhiqi Shen are with the College of Computing and Data Science, Nanyang Technological University, Singapore. (E-mail: zhan0552@e.ntu.edu.sg; \{xin.zhou, zhiwei.zeng, yongjie.wang, zqshen\}@ntu.edu.sg).}
\thanks{Manuscript received April 19, 2021; revised August 16, 2021.}}

\markboth{Journal of \LaTeX\ Class Files,~Vol.~14, No.~8, August~2021}%
{Shell \MakeLowercase{\textit{et al.}}: A Sample Article Using IEEEtran.cls for IEEE Journals}


\maketitle

\begin{abstract}

MultiModal Recommendation (MMR) systems have emerged as a promising solution for improving recommendation quality by leveraging rich item-side modality information, prompting a surge of diverse methods.
Despite these advances, existing methods still face two critical limitations. First, they use raw modality features to construct item–item links for enriching the behavior graph, while giving limited attention to balancing collaborative and modality-aware semantics or mitigating modality noise in the process. Second, they use a uniform alignment weight across all entities and also maintain a fixed alignment strength throughout training, limiting the effectiveness of modality–behavior alignment. To address these challenges, we propose \textbf{EGRA}. First, instead of relying on raw modality features, it alleviates sparsity by incorporating into the behavior graph an item–item graph built from representations generated by a pretrained MMR model. This enables the graph to capture both collaborative patterns and modality-aware similarities with enhanced robustness against modality noise. Moreover, it introduces a novel bi-level dynamic alignment weighting mechanism to improve modality-behavior representation alignment, which dynamically assigns alignment strength across entities according to their alignment degree, while gradually increasing the overall alignment intensity throughout training. Extensive experiments on five datasets show that EGRA significantly outperforms recent methods, confirming its effectiveness.


\end{abstract}

\section{Introduction}\label{introduction}
MultiModal Recommendation systems have emerged as a promising solution for enhancing recommendation quality by incorporating rich modality information from items. By leveraging multiple sources of information, they can more accurately model user preferences, leading to substantial performance gains over traditional methods \cite{liu2022multimodal, xu2021multi, bu2010music, baltruvsaitis2018multimodal, zhou2025learning, zhang2025semantic}.

The majority of MMRs adopt graph-based designs, applying graph neural networks to learn from the associations between user–item interactions and item modality features. A prevailing paradigm for leveraging item modality feature is to construct an item–item semantic similarity graph based on multimodal features, and then to apply graph convolution separately on the item–item semantic graph and the user–item behavior graph to learn semantic and behavioral representations, which often leads to improved performance \cite{lattice, freedom, micro}. In addition, several methods further address modality noise \cite{xv2024improving, mgcn}, enhance the learning of user-side modality features \cite{gume} or model user preferences over specific modalities \cite{dualgcn, mgcn}. Despite these advances, they still suffer from two critical limitations.

First, existing methods pay limited attention to balancing collaborative and modality-aware semantics, as well as mitigating the impact of modality noise when addressing behavior graph sparsity. Recent approaches, such as~\cite{gume}, typically enhance the behavior graph by incorporating item–item links derived solely from raw modality feature similarities. While this strategy enriches the structure, the resulting links primarily capture superficial modality resemblance, with collaborative signals only indirectly modeled through user nodes—potentially biasing representation learning. For example, a table tennis paddle may be linked to other paddles with similar appearances, rather than to table tennis balls that are more likely to be co-purchased. Such modality-driven links emphasize surface-level resemblance while overlooking co-purchase patterns, causing the model to overfit to modality cues, underrepresent collaborative intent, and ultimately produce biased item representations. Furthermore, noisy or low-quality modality features may introduce spurious item–item links, degrading graph quality~\cite{xv2024improving}. Therefore, it is essential to enrich the behavior graph with item–item links that jointly reflect collaborative and modality-aware semantics, while remaining robust to modality noise.

Second, existing methods often have limited capacity to effectively align modality and behavior representations. They typically apply a uniform alignment strength across all entities (i.e., users and items), and maintain this fixed value throughout training \cite{freedom, mgcn, gume}. This overlooks two fundamental issues. On one hand, different entities may exhibit varying degrees of discrepancy between their modality and behavior representations, necessitating different alignment strengths. For example, frequently interacted entities are sampled more often during training and thus receive repeated alignment updates, whereas sparsely interacted ones appear infrequently and remain weakly aligned. On the other hand, applying a constant alignment strength across epochs fails to reflect the changing stability of representations over time. Strong alignment imposed too early—when behavior and modality embeddings are still unstable—can interfere with the learning of core behavioral patterns and lead to suboptimal convergence. Therefore, it is necessary to adaptively adjust alignment strength both across entities and over training epochs, enabling more personalized and progressive representation alignment.

To tackle these challenges, we propose a novel multimodal recommendation framework: ``Toward \textbf{E}nhanced Behavior \textbf{G}raphs and \textbf{R}epresentation \textbf{A}lignment for Multimodal Recommendation'' (EGRA), which enhances behavior graphs and representation alignment to improve recommendation performance. Specifically, EGRA mitigates behavior graph sparsity by incorporating an item–item graph constructed using representations generated by a pretrained MMR model. It comprehensively models both collaborative and modality-aware similarities, while being less affected by noise in raw modality features. Additionally, EGRA incorporates a bi-level dynamic alignment weighting mechanism to enhance modality-behavior representation alignment. It adaptively adjusts alignment strength in an entity-wise manner based on the alignment degree, and also progressively increases the overall strength in epoch-wise manner to stabilize early training and improve alignment consistency over time.
Extensive experiments on five public benchmark datasets show that EGRA consistently outperforms state-of-the-art multimodal recommendation methods, validating its effectiveness.

The contributions of this paper are summarized as follows:
\begin{itemize}
    \item We propose a behavior graph enhancement method to improve behavior representations, which augments the behavior graph with an item–item graph built from pretrained representations, capturing both collaborative and modality-aware similarities while reducing the impact of modality noise.

    \item We propose a bi-level dynamic alignment weighting mechanism that adjusts alignment strength in both entity-aware and epoch-aware manners to enhance modality-behavior representation alignment.

    \item We conduct extensive experiments on five public datasets, showing that EGRA consistently outperforms state-of-the-art MMR methods, confirming its effectiveness.
\end{itemize}

\section{Related Work}
\subsection{Multimodal Recommendation}

Compared with earlier MMR methods \cite{vbpr, dvbpr, liu2017deepstyle, vecf}, GCN-based approaches have demonstrated superior effectiveness and gained increasing attention in recent research.

MMGCN \cite{mmgcn} applies multiple GCN modules over behavior graph to learn modality-specific features for users and items, which are subsequently fused to form comprehensive representations. GRCN \cite{grcn} adaptively refines the user-item interaction graph structure to alleviate noise caused by implicit feedback. BM3 \cite{bm3} explores the application of self-supervised learning within MMR by employing embedding dropout to construct contrastive views, eliminating the reliance on negative samples and auxiliary graphs and achieving higher computational efficiency. Besides, hypergraph structures for modeling global user interests are also explored in LGMRec~\cite{lgmrec}. 

However, these methods overlook the diversity in user preferences across modalities. DualGNN \cite{dualgcn} tackles this problem by constructing a user co-occurrence graph, where modality embeddings are weighted and used as the initial node features. Through graph convolution on the co-occurrence graph, user modality preferences are implicitly captured. MGCN \cite{mgcn} models users' modality preferences by analyzing their interaction behavior features, and subsequently adapts the integration of item modality features based on the inferred preferences. 

LATTICE \cite{lattice} and MICRO \cite{micro} aim to uncover the underlying semantic relationships between items by utilizing multimodal features and employ graph convolution to inject item modality similarities into item representations. Nevertheless, both incur high computational costs from dynamically updating item semantic graphs during training. To mitigate this, FREEDOM \cite{freedom} shows that precomputing and freezing the semantic graph before training can still yield strong performance. DA-MRS \cite{xv2024improving} further addresses noise in user behaviors and multimodal content by estimating interaction reliability via multimodal signals and adjusting the BPR loss. These methods typically model user modality features by aggregating the modality features of interacted items. In contrast, GUME \cite{gume} enhances user modality representations by performing graph convolution over the user–item graph with item modality features as input, followed by refining the learned representations via mutual information maximization. It also constructs an item–item graph by identifying common neighbors across different modalities and uses it to alleviate behavior graph sparsity. 

\vspace{-0.95em}
\subsection{Long-tail Recommendation}
Under large-scale item collections, the long-tail effect intensifies in recommendation systems, often compromising recommendation accuracy \cite{gume, yin2012challenging, volkovs2017dropoutnet}. To mitigate these issues, the paper\cite{yin2012challenging} applies random walk-based strategies on the interaction graph to uncover underexposed items that match user preferences. DropoutNet \cite{volkovs2017dropoutnet}, on the other hand, applies dropout-based training to leverage content features for predictions, particularly when historical data is sparse for certain users or items. Inspired by meta-learning, MeLU \cite{lee2019melu} tackles the user cold-start challenge by leveraging meta-learning principles, employing a model-agnostic strategy to quickly adapt to new users with minimal data. Other lines of research focus on enhancing the exposure and ranking of long-tail items. TailNet \cite{liu2020long} proposes a session-aware preference adjustment mechanism to dynamically balance recommendations between popular and less-interacted items. MIRec \cite{zhang2021model} presents a dual transfer learning approach designed to bridge the gap between popular and less popular items by transferring knowledge from the former to the latter. GALORE \cite{luo2023improving} and GraphDA \cite{fan2023graph} enhance the structural connectivity of tail items by strengthening the links between head and tail items, while intentionally reducing the dominance of head items through selective removal of their links to users. DiffuASR \cite{liu2023diffusion} applies a diffusion-based sequence generation framework to create high-quality pseudo interactions that mitigate data sparsity and long-tail user issues in sequential recommendation. LinRec \cite{liu2023linrec} introduces an efficient L2-normalized linear attention mechanism for Transformer-based sequential recommendation, achieving comparable or better performance than standard attention while substantially reducing computational and memory costs for long sequences. Besides, CoRAL\cite{wu2024coral}, LLM-ESR \cite{liu2024llm} and Llama4Rec \cite{luo2024integrating} explore how to use LLM to enhance performance on long-tail items.
\begin{figure*}[h]
    \centering
\includegraphics[width=0.98\textwidth]{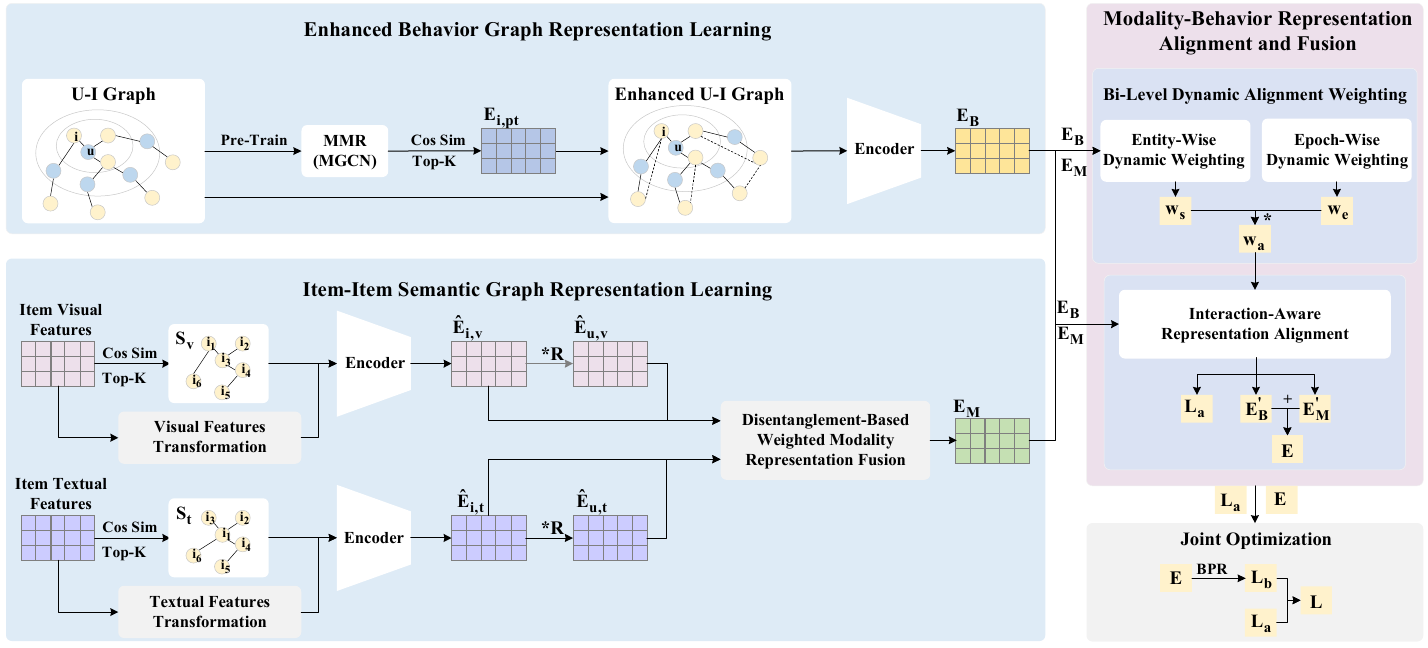} 
    \caption{The overall architecture of \textbf{EGRA}. It involves three graphs: an enhanced U-I graph with item–item relations from a pretrained MGCN, and two modality-specific semantic graphs. After encoding behavior and modality representations on these graph, it dynamically adjusts alignment strength in both entity-wise and epoch-wise manner to generate personalized alignment weights with the proposed Bi-Level Dynamic Alignment Weighting. It further performs Interaction-Aware Representation Alignment, with interaction context being another guidance signal.}   
    \label{fig:egra}  
\end{figure*}

\section{Preliminaries}\label{pf}

Let $U = \{u\}$ and $I = \{i\}$ denote the sets of users and items, respectively. The interaction behaviors between users and items are captured by a binary matrix $\mathbf{R} \in \{0,1\}^{|U| \times |I|}$, where an entry $R_{u,i} = 1$ signifies that user $u$ has interacted with item $i$, and $R_{u,i} = 0$ otherwise. This interaction matrix can also be represented as a bipartite graph $\mathcal{G} = (V, \mathcal{E})$, where the node set is defined as $V = U \cup I$, and the edge set $\mathcal{E} = \{(u,i) \mid R_{u,i} = 1, u \in U, i \in I\}$ encodes the interactions.

Each user and item is associated with a distinct identifier embedding, and all embeddings are jointly stored in a matrix $\mathbf{E}_{B} \in \mathbb{R}^{(|U| + |I|) \times d}$, where $d$ indicates the dimension of the embedding space. For each modality $m \in M$, the features of all items are collected into a matrix $\mathbf{E}_m \in \mathbb{R}^{|I| \times d_m}$, with $d_m$ being the dimension of modality $m$. In this work, we focus on \textbf{v}isual and \textbf{t}extual modality, and thus $M = \{v, t\}$. We denote the feature of item $i$ in modality $m$ by $\mathbf{E}_{i,m}$.


\section{Methodology}
This section provides a comprehensive overview of the proposed \textbf{EGRA}, covering both its overall architecture and the design of its key components. The full structure is illustrated in Figure~\ref{fig:egra}, and each module is described in the following.


\subsection{Enhanced Behavior Graph Representation Learning}
This component aims to mitigate sparsity of behavior graph and further enhances user and item representations.  
\subsubsection{Enhanced Behavior Graph Construction}
As noted in Section~\ref{introduction}, existing strategies often overemphasize modality features while overlooking collaborative signals as well as noise in raw modality features. To address these, we adopt a pre-training strategy \cite{fan2023graph}, where a lightweight multimodal model is first pretrained to generate item representations. The representations are then used to construct a more accurate and semantically rich item–item graph, which is integrated to enhance the behavior graph.

In this work, we adopt the model \textbf{MGCN} \cite{mgcn} as the pretraining backbone, motivated by the following considerations: (1) MGCN aims to reduce modality noise by leveraging behavior representations and demonstrates its effectiveness in suppressing the adverse effects of noisy modality features. As a result, the learned representations are less affected by modality noise. (2) MGCN aligns behavior and modality representations and integrates them to produce final item embeddings that encode both collaborative signals and modality similarity information across items. As a result, the item-item graph constructed from these embeddings is more accurate and comprehensive than those based solely on raw modality features. (3) MGCN is a lightweight model with a modest number of parameters and requires tuning only a single hyperparameter with three candidate values, while still achieving competitive performance. Owing to its efficiency and effectiveness, it is often adopted as the foundational backbone for subsequent models, including the state-of-the-art GUME \cite{gume}. (4) Moreover, the experimental results in Section \ref{spm} indicate that, compared with using the strongest-performing but more complex model as the pre-training backbone, adopting MGCN has achieved competitive performance.

With item embeddings matrix $\mathbf{E}_{pt}$ obtained from the pre-trained MGCN, we construct the item-item semantic graph $\mathbf{S}^{pt}$ by computing pairwise cosine similarity and applying a $\textbf{Top-K}$ nearest neighbor selection strategy \cite{fedc, chen2009fast}.
Concretely, for a given item pair $(i, j)$, the semantic similarity score $\mathbf{S}^{pt}_{i,j}$ is computed as the cosine similarity between their respective pre-trained representations (Namely, $\mathbf{E}_{i,pt}$ and $\mathbf{E}_{j,pt}$), formally: 

\begin{equation} 
    \mathbf{S}^{pt}_{i,j} = \text{Cos}(\mathbf{E}_{i,pt}, \mathbf{E}_{j,pt}). 
\end{equation}

To emphasize the most relevant semantic relations, we retain the nearest $H$ neighbors for each item based on their $\text{S}^{pt}_{i,j}$ scores. Furthermore, for simplicity, the weight of each retained edge is set as 1. Formally, the edge weight $\mathbf{S}^{pt}_{i,j}$ is updated as: 

\begin{equation}\label{sm}
    \mathbf{S}^{pt}_{i,j} = \left\{
\begin{array}{ll}
1, & \mathbf{S}^{pt}_{i,j} \in \text{Top-H}({\mathbf{S}^{pt}_{i,k}~|~k \in I}). \\
0, & \text{otherwise. }
\end{array}
\right.
\end{equation}

Then, we symmetrize the interaction graph $\textbf{R}$ and incorporate edges from the item-item semantic graph $\mathbf{S}^{pt}$, yielding the enhanced behavior graph as follows:
\[
     \mathbf{G}_e = 
 \begin{pmatrix}
 0 & \mathbf{R}\\
 \mathbf{R}^T & \mathbf{S}^{pt}
 \end{pmatrix}.
\]

\subsubsection{Enhanced Behavior Graph Encoding}

We apply the LightGCN \cite{lightgcn} to encode user and item ID embedding on $\mathbf{G}_e$. The embedding propagation at the $l$-th layer is:
 \begin{equation}
     \mathbf{E}^l_{B} = (\mathbf{D}^{-\frac{1}{2}}_{\mathbf{G}_e} \mathbf{G}_e\mathbf{D}^{-\frac{1}{2}}_{\mathbf{G}_e}) \mathbf{E}^{l-1}_{B}
 \end{equation}
where $\mathbf{E}^l_{B}$ is the user and item ID embedding matrix at the $l$-th layer and $D_{\mathbf{G}_e}$ is the diagonal degree matrix of ${\mathbf{G}_e}$, used for symmetric normalization. 

The final ID embedding is computed as the mean of all layer-wise representations:
\begin{equation}
    \mathbf{E}_{B} = \frac{1}{L+1}\sum^L_{l=0} \mathbf{E}^l_{B}
\end{equation}

\subsection{Item-Item Semantic Graph Representation Learning}
This component aims to refine modality-specific item representations by capturing fine-grained semantics.
\subsubsection{Item-Item Semantic Graph Construction}

In line with established practices \cite{mgcn, gume}, we construct the modality-specific item-item semantic graph $\mathbf{S}_m$ for each modality $m$ by utilizing raw modality features and applying a $\textbf{Top-K}$ neighbor selection strategy, which is also similar to the construction of $\mathbf{S}^{pt}$. The only distinction is that each retained edge preserves its original cosine similarity as the edge weight.

The collection of weights $\text{S}^m_{i,j}$ across all item pairs $(i, j)$ constitutes the item-item semantic graph for modality $m$. To stabilize training and mitigate potential gradient explosion, we further normalize the item-item semantic graph matrix $\textbf{S}^m$ as: 
\begin{equation} 
    \hat{\mathbf{S}}^m = \mathbf{D}^{-\frac{1}{2}}_{\mathbf{S}^m} {\mathbf{S}^m} \mathbf{D}^{-\frac{1}{2}}_{\mathbf{S}^m} \end{equation} 
where $\mathbf{D}_{\mathbf{S}^m}$ is the diagonal degree matrix associated with $\mathbf{S}^m$.

\subsubsection{Item-Item Semantic Graph Encoding}

Given the inconsistency in dimensionality between raw modality features and ID embeddings, we employ a two-layer MLP transformation function to align modality features space with that of ID representations, following \cite{mgcn,gume}. Formally, the mapping is: 
\begin{equation}\label{fm} 
f_m(\mathbf{E}) = \sigma(\mathbf{W}^2_m(\mathbf{W}^1_m\mathbf{E}^T + \mathbf{b}^1_m)+\mathbf{b}^2_m) 
\end{equation} 
where ``$T$'' is matrix transposition; $\mathbf{W}^{1}_m \in \mathbb{R}^{d \times d_m}$, $\mathbf{W}^{2}_m \in \mathbb{R}^{d \times d}$ and $\mathbf{b}^{1/2}_m \in \mathbb{R}^d$ are the modality-specific transformation parameters; $\sigma$ represents the Sigmoid activation function.

To mitigate the potential noise present in raw modality features, we enhance the transformed modality representation by incorporating the corresponding behavior-based embedding as a corrective signal. This adjustment is formally defined as: 
\begin{equation}\label{purify} 
\hat{\mathbf{E}}_{i,m} = f_m(\mathbf{E}_{i,m}) \odot \mathbf{E}_{i, B} 
\end{equation} 
where ``$\odot$'' denotes the Hadamard product.

Then, we refine modality-specific semantic representations for items by propagating information over the semantic graph $\hat{\mathbf{S}}^m$ with graph convolution. The $l-$th layer propagation is: 

\begin{equation} 
\hat{\mathbf{E}}^l_{i,m} = \hat{\mathbf{S}}^m \hat{\mathbf{E}}^{l-1}_{i,m} 
\end{equation} 
where the initial (i.e., the 0-th layer) embeddings correspond to the corrective modality features from Equation \eqref{purify}.

The final-layer output, abbreviated as $\hat{\mathbf{E}}_{i,m}$, is utilized as the semantic representation of item $i$ under modality $m$. We derive the users' semantic representations by aggregating those of their interacted items, as follows:
\begin{equation}\label{ueu}
    \hat{\mathbf{E}}_{u,m} = (\mathbf{D}^{-\frac{1}{2}}_\mathbf{R} \mathbf{R}\mathbf{D}^{-\frac{1}{2}}_\mathbf{R}) \hat{\mathbf{E}}_{i,m}~
\end{equation}
where $\mathbf{D}_\mathbf{R}$, the diagonal degree matrix of interaction graph $\mathbf{R}$, is used for symmetric normalization.

Finally, we concatenate the learned user and item embeddings for modality $m$ into $\hat{\mathbf{E}}_m \in \mathbb{R}^{(|U| + |I|) \times d}$.

\subsection{Representation Fusion}


\subsubsection{Modality Representation Fusion}
Different modalities may contain overlapping semantic information~\cite{mgcn}, while users often exhibit varying preferences across modalities~\cite{dualgcn}. Considering it, we apply a Disentanglement-based Weighted Modality Representation Fusion to fuse modality representations. It first separates modality-specific and shared components and then performs preference-guided weighted fusion.

Specifically, we extract the shared content across modalities using an attention mechanism. The attention function for the representation of modality $m$ (i.e., $\hat{\mathbf{E}}_m$) is defined as:  
\begin{equation}
    f_{att}(\hat{\mathbf{E}}_m) = tanh(\hat{\mathbf{E}}_m\mathbf{W}_a + \mathbf{b}_a)\mathbf{V}^T_a
\end{equation}
where ``$T$'' is matrix transposition; $\mathbf{V}_a \in \mathbb{R}^d$, $\mathbf{W}_a \in \mathbb{R}^{d\times d}$ and $\mathbf{b}_a \in \mathbb{R}^d$
are attention parameters shared across all modalities.

We apply softmax to normalize attention weights and compute a weighted fusion of modality-specific features as the shared modality representation, formulated as:
\begin{equation}
    \mathbf{E}_s = \sum_{m\in M} \frac{exp(f_{att}(\hat{\mathbf{E}}_m))}{\sum_{m'\in M}exp(f_{att}(\hat{\mathbf{E}}_{m'}))} \hat{\mathbf{E}}_m
\end{equation}

The exclusive feature for modality $m$ is obtained by subtracting the shared representation $\mathbf{E}_s$ from its modality feature $\hat{\mathbf{E}}_m$. The user's preference for the exclusive feature is measured based on the behavior representations, as follows:
\begin{equation}
    \mathbf{P}_m = \sigma( \mathbf{E}_{B} \mathbf{W}_{p,m} + \mathbf{b}_{p,m})
\end{equation}
where $\mathbf{W}_{p,m}\in \mathbb{R}^{d\times d}$ and $\mathbf{b}_{p,m} \in \mathbb{R}^d$ are modality-specific.

The final fused modality representation is computed as:
\begin{equation}
    \mathbf{E}_{M} = \frac{1}{|M|+1}  (\mathbf{E}_s + \sum_{m\in M} \mathbf{P}_m \odot (\hat{\mathbf{E}}_m - \mathbf{E}_s) )
\end{equation}

\subsubsection{Modality-Behavior Representation Alignment and Fusion}
We first align the modality and behavior representations, then fuse them to obtain the final embeddings. 

To enhance representation alignment and better balance the contributions of the main loss and alignment loss, we propose the \textbf{Bi-Level Dynamic Alignment Weighting} mechanism. It enables personalized and progressive control over alignment strength for both users and items by dynamically adjusting it in an entity-wise and epoch-wise manner during training. 

$\noindent \bullet$ \textbf{Entity-Wise Dynamic Weighting} within each batch. For each mini-batch, we evaluate the alignment degree between behavior and modality representations for every user and item. Based on this, we assign entity-specific alignment weights—giving higher weights to poorly aligned entities to encourage stronger alignment, while assigning lower weights to well-aligned ones. Formally, take the user $u$ in batch $\mathcal{B}$ as an example, the entity-wise dynamic weighting within each batch is:

\begin{equation}
    \omega^{\mathcal{B}}_{u,s} = \frac{exp(\frac{1-S(e^B_{u}, e^M_{u})}{\tau_1})}{\sum_{v \in \mathcal{B}} exp(\frac{1-S(e^B_{v}, e^M_{v})}{\tau_1})} 
\end{equation}
where $S(\cdot)$ denotes cosine similarity; $e^B_u$ and $e^M_u$ are user $u$'s behavior and modality representation.

$\noindent \bullet$ \textbf{Epoch-Wise Dynamic Weighting}. To stabilize training and prevent premature dominance of the alignment loss, we start with a small alignment weight and gradually increase it over training epochs. The weight is eventually fixed after reaching a predefined upper bound. Formally, taking user $u$ in batch $B$ at the $p$-{th} epoch as an example, the alignment weight for user $u$ is defined as: 
\vspace{-1pt}
\begin{equation}
    \omega^p_{u, e} =
    \begin{cases}
        (\lambda_{min} + \frac{p}{P}\times(\lambda_{max} - \lambda_{min})),  &p < P. \\
        &\\
        \lambda_{max}, & p \ge P.
    \end{cases}
\end{equation}

\noindent where $\lambda_{min}$ and $\lambda_{max}$ denote the initial and stable epoch-wise alignment weights, respectively; $p$ is the current epoch and $P$ is the number of warm up epoch from which the epoch-wise weight remains fixed.

The total alignment weight assigned to user $u$ in batch $\mathcal{B}$ at epoch $p$ is defined as:
\begin{equation}
    \omega_{u,a}(p, \mathcal{B}) = \omega^p_{u,e} \times \omega^{\mathcal{B}}_{u,s}
\end{equation}

Next, we adopt contrastive learning to align modality and behavior representations. Specifically, we design the Interaction-Aware Representation Alignment mechanism, which is not merely enforced in the geometric embedding space, but is further guided by semantic grounding. It aims to enhance consistency between modality-driven and behavior-driven representations while preventing the user–item interaction structure encoded in the representations from being disrupted during the alignment process. To this end, on the one hand, it directly pulls the two types of representations closer; on the other hand, it further enhances alignment by leveraging interaction context as an anchor, where for each $(u, i)$ pair in the current batch, the interacted item $i$ serves as an anchor to guide the alignment between user $u$'s modality and behavior representations. Formally, the context-aware representation alignment for user $u$ is defined as follows:
{\small
\begin{equation}
    \mathcal{L}^u_{a} = -log \left( \frac{exp(\frac{S(e^B_u, e^M_u)}{\tau_2})  + \varphi(e^B_u, e^M_u, e^f_i|\tau_3) } {
    \sum_{v\in \mathcal{B}_u} exp(\frac{S(e^B_u, e^M_v)}{\tau_2}) + \sum_{j \in \mathcal{B}_i}\varphi(e^B_u, e^M_u, e^f_j|\tau_3) }
    \right)
\end{equation}
}
\noindent where $S(\cdot)$ denotes cosine similarity; $e^f_i = \frac{e^B_i + e^M_i}{2}$; $\mathcal{B}_u$ and $\mathcal{B}_i$ denote the user and item sets in the current batch, respectively; $\varphi(e_1, e_2, e_3|\tau)$ is defined as follows:
\begin{equation}
    \varphi(e_1, e_2, e_3|\tau) =  exp(\frac{S(e_1, e_3)}{\tau}) + exp(\frac{S(e_2, e_3)}{\tau})
\end{equation}

With the personalized weight as a guidance signal, the representation alignment loss for users in batch $\mathcal{B}$ at epoch $p$ is computed as:
\begin{equation}
    \mathcal{L}^{u}_a(p, \mathcal{B}) = \sum_{u\in \mathcal{B}} \omega_{u,a}(p, \mathcal{B}) \times \mathcal{L}^u_a
\end{equation}

Similarly, the representation alignment loss for items in batch $\mathcal{B}$ at epoch $p$ is denoted as $\mathcal{L}^{i}_a(p, \mathcal{B})$.

The total alignment loss for all users and items across batches at epoch $p$, denoted as $\mathcal{L}_a$, is defined as:
\begin{equation}
    \mathcal{L}_a = \sum_{\mathcal{B}}(\mathcal{L}^{u}_a(p, \mathcal{B}) + \mathcal{L}^{i}_a(p, \mathcal{B}))
\end{equation}

After performing representation alignment, we obtain the final user and item representations by summing their behavior and modality representations. Formally:
\begin{equation}
    \mathbf{E} = \mathbf{E}_{B}+ \mathbf{E}_M
\end{equation}

\vspace{-22pt}

\subsection{Model Optimization}

We define the predicted interaction score $\hat{y}_{ui}$ as the inner product between the user representation $\mathbf{E}_u$ and the item representation $\mathbf{E}_i$, formulated as: 
\begin{equation} 
    \hat{y}_{ui} = \mathbf{E}_{u} \cdot \mathbf{E}_{i} 
\end{equation}

We adopt the BPR loss as the primary optimization objective. The loss function is defined as:
\begin{equation}
    \mathcal{L}_{b} = \sum_{(u,i,j) \in D} -log \sigma(\hat{y}_{ui} - \hat{y}_{uj})
\end{equation}
where $D$ is training set; $(i, j)$ represents a pair of interacted and non-interacted items for user $u$.

The final optimization objective jointly integrates the BPR loss, the proposed modality-behavior alignment loss, and an $L_2$ regularization term applied to the embeddings to prevent overfitting, which is formally expressed as: 
\begin{equation} 
    \mathcal{L} = \mathcal{L}_{b} + \mathcal{L}_a + \lambda_2||\mathbf{E}||_2 
\end{equation}

\section{Experiments}
To assess EGRA, we conduct extensive experiments on five datasets to address the following research questions:

\noindent \textbf{RQ1:} How significantly does EGRA improve recommendation accuracy over SOTA multimodal approaches?

\noindent \textbf{RQ2:} How does the proposed behavior graph augmentation contribute to enhancing the performance for long-tail items?

\noindent \textbf{RQ3:} How do the critical components of EGRA contribute to the improvement of recommendation performance?

\noindent \textbf{RQ4:} How does the proposed behavior graph augmentation outperform existing multimodal-based augmentation methods? 

\noindent \textbf{RQ5:} How do different pre-training backbones affect the performance of EGRA? 

\noindent \textbf{RQ6:} How does EGRA perform in terms of training efficiency compared with recent models?

\noindent \textbf{RQ7:} How does EGRA respond to changes in hyperparameter configurations in terms of performance? 
\vspace{-8pt} 
\subsection{Experimental Configuration}
\noindent \textbf{Dataset.} Following established practices~\cite{mgcn,lattice,lgmrec}, we conduct experiments on five benchmark datasets: Baby, Sports, Clothing, MicroLens~\cite{ni2023content} and Elec.

\begin{table*}[t]
    \def\arraystretch{1.25}	
    \caption{Performance comparison of various recommendation models. The best-performing results are highlighted in \textbf{bold}, and the second-best results are indicated with \underline{underline}. The \textbf{Imp.} column indicates the relative performance gain over the strongest baseline.}
    \label{tab:me}
    \centering
    \tabcolsep=0.1cm
    \renewcommand{\arraystretch}{0.8}
    \begin{tabular}{lcccccccccccccc}
        \toprule
        \multirow{2}{*}{Datasets}& \multirow{2}{*}{Metrics} & \textbf{LightGCN} && \textbf{VBPR} & \textbf{LATTICE} &\textbf{SLMRec} &\textbf{FREEDOM} &\textbf{BM3} & \textbf{MGCN} & \textbf{LGMRec} & \textbf{DA-MRS} & \textbf{GUME} & \textbf{EGRA} & \textbf{Imp.}\\
        \cline{3-3}
        \cline{5-15}
                                                          & &SIGIR‘20 && AAAI’16 & MM‘21 & TMM’22 &MM'23 & WWW‘23 & MM'23& AAAI’24 & KDD‘24 &CIKM'24 &Ours & \\
        \midrule
        \multirow{4}{*}{Baby}     &R@10&0.0479 &&0.0423 &0.0547 &0.0547 &0.0624 &0.0542 &0.0607 &0.0645 &0.0650 &\underline{0.0683} &\textbf{0.0703} &2.93\%\\
                                  &R@20 &0.0754 &&0.0664 &0.0844 &0.0810 &0.0985 &0.0862 &0.0950 &0.0981 &0.0994 &\underline{0.1039} &\textbf{0.1058} &1.83\%\\
                                  &N@10  &0.0257 &&0.0223 &0.0289 &0.0285 &0.0324 &0.0285 &0.0328 &0.0350 &0.0346 &\underline{0.0369} &\textbf{0.0379} &2.71\%\\
                                  &N@20  &0.0328 &&0.0284 &0.0366 &0.0357 &0.0416 &0.0367 &0.0416 &0.0437 &0.0435 &\underline{0.0460} &\textbf{0.0470} &2.17\%\\
        \midrule

        \multirow{4}{*}{Sports}   &R@10&0.0569 &&0.0561 &0.0626 &0.0676 &0.0713 &0.0619 &0.0737 &0.0724 &0.0751 &\underline{0.0784} &\textbf{0.0819} &4.46\%\\
                                  &R@20&0.0864 &&0.0859 &0.0959 &0.1017 &0.1077 &0.0971 &0.1107 &0.1087 &0.1125 &\underline{0.1167} &\textbf{0.1213} &3.94\%\\
                                  &N@10  &0.0311 &&0.0307 &0.0337 &0.0374 &0.0382 &0.0338 &0.0403 &0.0392 &0.0402 &\underline{0.0428} &\textbf{0.0449} &4.91\%\\
                                  &N@20  &0.0387 &&0.0384 &0.0423 &0.0462 &0.0476 &0.0429 &0.0499 &0.0485 &0.0498 &\underline{0.0527} &\textbf{0.0551} &4.55\%\\
        \midrule

        \multirow{4}{*}{Clothing} &R@10&0.0361 &&0.0283 &0.0468 &0.0540 &0.0624 &0.0425 &0.0658 &0.0549 &0.0647 &\underline{0.0695} &\textbf{0.0714} &2.73\% \\
                                  &R@20&0.0544 &&0.0417 &0.0688 &0.0810 &0.0928 &0.0637 &0.0963 &0.0822 &0.0963 &\underline{0.1008} &\textbf{0.1043} &3.47\%\\
                                  &N@10  &0.0197 &&0.0157 &0.0256 &0.0285 &0.0336 &0.0232 &0.0359 &0.0301 &0.0353 &\underline{0.0378} &\textbf{0.0390} &3.17\%\\
                                  &N@20  &0.0243 &&0.0191 &0.0312 &0.0357 &0.0414 &0.0286 &0.0436 &0.0370 &0.0433 &\underline{0.0458} &\textbf{0.0473} &3.28\%\\

        \midrule

        \multirow{4}{*}{MicroLens} &R@10&0.0720 &&0.0677 &0.0726 &0.0778 &0.0674 &0.0606 &0.0756 &0.0748 &\underline{0.0815} &0.0813 &\textbf{0.0890} &9.20\%\\
                                  &R@20&0.1075 &&0.1026 &0.1089 &0.1190 &0.1032 &0.0981 &0.1134 &0.1132 &\underline{0.1221} &0.1207 &\textbf{0.1302} &6.63\%\\
                                  &N@10  &0.0376 &&0.0351 &0.0380 &0.0405 &0.0345 &0.0304 &0.0387 &0.0390&\underline{0.0431} &0.0424 &\textbf{0.0469} &8.82\%\\
                                  &N@20  &0.0467 &&0.0441 &0.0473 &0.0511 &0.0437 &0.0400 &0.0484 &0.0489 &\underline{0.0536} &0.0525 &\textbf{0.0575} &7.28\%\\

        \midrule

        \multirow{4}{*}{Elec} &R@10&0.0363 &&0.0293 &- &0.0422 &0.0396 &0.0437 &0.0442 &0.0417 &0.0435 &\underline{0.0458} &\textbf{0.0498} &8.73\%\\
                                  &R@20&0.0540 &&0.0453 &- &0.0630 &0.0601 &0.0648 &0.0650 &0.0625 &0.0644 &\underline{0.0680} &\textbf{0.0732} &6.63\%\\
                                  &N@10  &0.0204 &&0.0159 &- &0.0237 &0.0220 &0.0247 &0.0246 &0.0233&0.0245 &\underline{0.0253} &\textbf{0.0282} &7.65\%\\
                                  &N@20  &0.0250 &&0.0202 &- &0.0291 &0.0273 &0.0302 &0.0302 &0.0287 &0.0299 &\underline{0.0310} &\textbf{0.0343} &10.65\%\\
        
        \bottomrule
    \end{tabular}
    \begin{tablenotes}
        \item[*] Note: ‘–’ indicates that the model runs out of memory on an A100 GPU with 80GB.
    \end{tablenotes}
    \vspace{-8pt}
\end{table*}

\noindent \textbf{Baselines.} 
We compare our model against widely used and state-of-the-art baselines: \textbf{LightGCN}, \textbf{VBPR} \cite{vbpr}, \textbf{LATTICE} \cite{lattice},  \textbf{SLMRec} \cite{slmrec}, \textbf{FREEDOM} \cite{freedom}, \textbf{MGCN} \cite{mgcn}, \textbf{BM3} \cite{bm3}, \textbf{LGMRec} \cite{lgmrec}, \textbf{DA-MRS} \cite{xv2024improving} and \textbf{GUME} \cite{lin2024gume}.

\noindent \textbf{Evaluation Protocols.} 
To comprehensively assess EGRA, we conduct both general and long-tail evaluations using two standard metrics: Recall@K (R@K) and NDCG@K (N@K), where K=10 and K=20. The evaluation metrics are obtained by averaging the results across all users in the testing set. Model training is terminated early if Recall@20 does not improve for 20 consecutive epochs. For the general evaluation, we adopt the data splitting strategy employed in prior works \cite{lattice,freedom}, where each user's interactions are randomly partitioned into training, validation, and testing sets following an 8:1:1 ratio.

\noindent \textbf{Implementation Specifications.} All baseline methods, as well as our EGRA, are implemented within the MMREC framework \cite{zhou2023mmrec}. The Top-K value used to construct the item-item semantic graph is uniformly set to 10 for all models. For optimization, we employ the Adam optimizer across all models. User and item embeddings are initialized using the Xavier initialization scheme, with the embedding dimensionality fixed at 64. For baseline models, we adopt the optimal hyperparameter configurations as reported in their original publications. In our model, the learning rate is fixed at $1 \times 10^{-3}$. The item neighbor number $H$ is fixed as 5 for all datasets. We apply three graph convolutional layers to the interaction graph, while the item-item semantic graph is encoded using two layers for the Baby dataset and a single layer for the remaining datasets. The remaining hyperparameters are selected through grid search: $\lambda_{min} \in [0, 0.005, 0.01]$; $\lambda_{max} \in \{0.02, 0.03, 0.04\}$; $P \in \{5, 10, 15, 20\}$; $\tau_1 \in \{0.6, 0.8, 1.0, 1.2, 1.4, 1.6, 1.8, 2.0\}$; $\tau_{2/3} \in \{0.1, 0.2, 0.3\}$.
\vspace{-8pt} 
\subsection{Performance Comparison}
\noindent \textbf{General Evaluation of \textbf{EGRA} (RQ1).} The performance comparison of EGRA with existing baselines under the general evaluation protocol is provided in Table \ref{tab:me}. 

From it, we can find that: EGRA demonstrates clear and consistent performance advantages across all datasets. Its effectiveness is particularly evident on the two larger datasets, Microlens and Elec, where it outperforms the most competing method by over 8.5\% on half of the evaluation metrics. The most significant improvement reaches as high as 10.65\%, indicating a substantial boost in recommendation quality. Even for the remaining metrics, the majority exhibit gains exceeding 7\%, with the smallest observed improvement still reaching 6.63\%, which remains close to that threshold, further underscoring EGRA's effectiveness. Among the remaining three datasets, EGRA delivers the strongest results on Sports, with all four metrics improving between 3.94\% and 4.91\%. On the Clothing dataset, EGRA achieves performance improvements with three metrics achieving gains above 3\%, and the remaining one metrics showing gains approaching 3\%. On the Baby dataset, two of the four metrics show improvements of nearly 3\%, while the remaining two maintain gains around 2\%. The results consistently demonstrate the effectiveness of EGRA.


\begin{figure}[h]
    \centering
\includegraphics[width=0.45\textwidth]{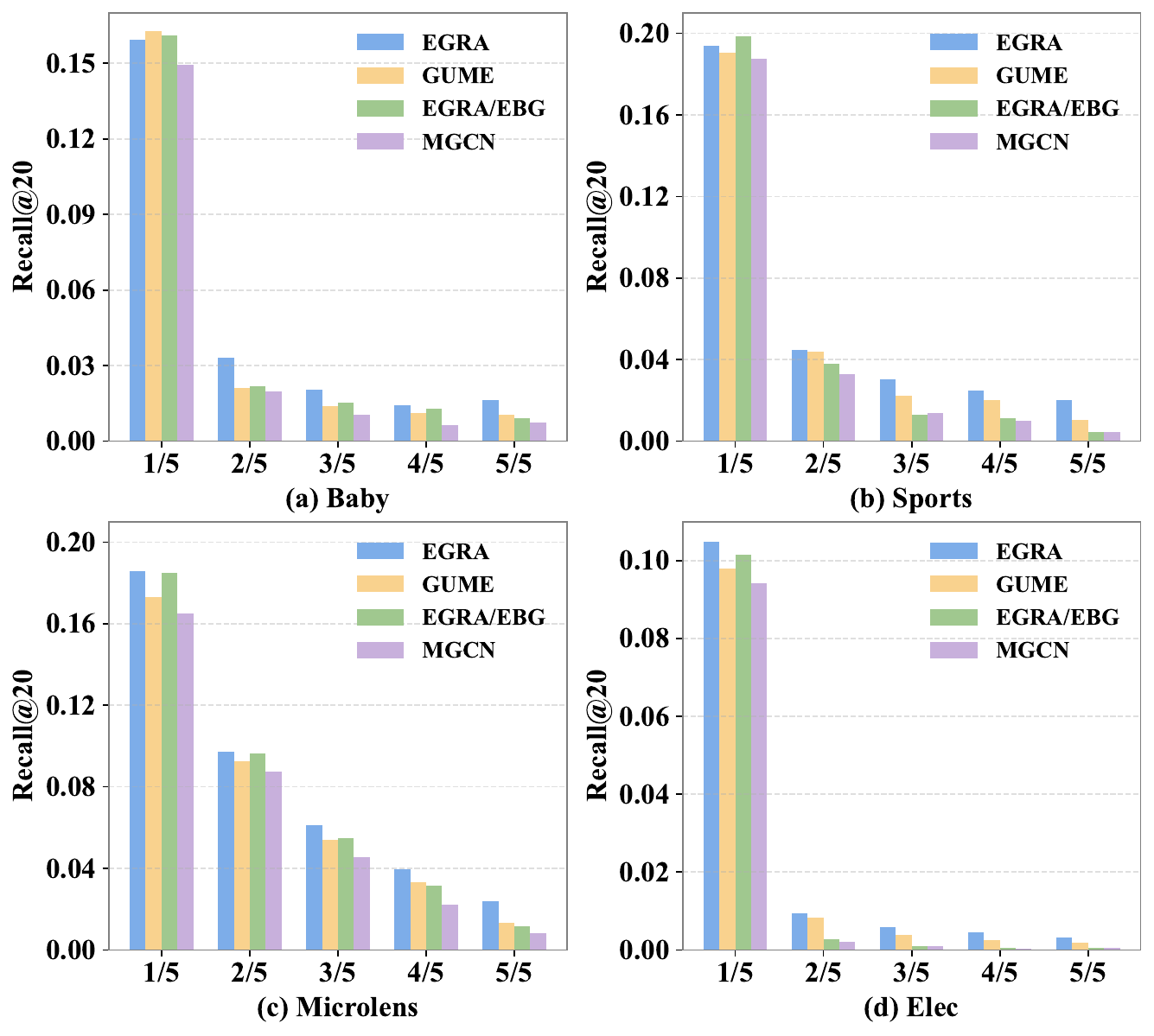} 
    \caption{Performance Analysis by Long-Tail Item Grouping.}  
    \label{fig:para_li}  
    \vspace{-20pt} 
\end{figure}

\noindent \textbf{Long-Tail Evaluation of EGRA (RQ2).} We further assess the effectiveness of EGRA in long-tail item scenarios by comparing it against recent baselines, including GUME, MGCN, and a variant of EGRA that replaces the enhanced behavior graph with the original behavior graph (\textbf{EGRA/EBG}). As interactions are often highly imbalanced, a small fraction of items typically dominates user engagement. Following this observation, the most active 20\% of items are categorized as head items, while the remaining 80\% are treated as tail items~\cite{gume}. Therefore, we sort items in the training dataset by interaction frequency in descending order and evenly partition them into five groups, where the $i$-th group ($i \in {1, 2, 3, 4, 5}$) corresponds to the top-$i/5$ segment of items. Based on this partition, we assign each user-item interaction in the testing set to the corresponding group. Then, the first group contains interactions involving head items, while the remaining four groups correspond to interactions with items at varying levels of long-tail sparsity. All models are trained following the same procedure as in the general evaluation setting. Once convergence is reached, we evaluate the performance on these five newly constructed test subsets. We perform experiments on four datasets, and the results are illustrated in Figure~\ref{fig:para_li}.

\textbf{First}, by comparing EGRA with EGRA/EBG, we observe that EGRA consistently outperforms EGRA/EBG across all four long-tail groups, with the performance advantage becoming more pronounced for items with higher degrees of tailness. This finding shows that the enhanced behavior graph enables EGRA to better alleviate the data sparsity of long-tail items, thereby yielding greater performance improvements.

\textbf{Second}, EGRA exhibits superior performance in addressing long-tail recommendation challenges compared to the baselines MGCN and GUME. In particular, EGRA consistently outperforms the recent best-performing GUME across all item groups and datasets. Notably, this includes the head item group, with the only exception observed in the Baby dataset. This shows that, compared to GUME, EGRA is more effective in improving performance for both long-tail and head items.

\vspace{-9.5pt} 
\subsection{Ablation Study (\textbf{RQ3})} \label{ab}

EGRA consists of two core components: the \textbf{E}nhanced \textbf{B}ehavior \textbf{G}raph (EBG) module and the \textbf{B}i-Level \textbf{D}ynamic \textbf{A}lignment Weighting (BDA) mechanism, where BDA includes \textbf{EN}tity-Wise Dynamic Weighting (EN) and \textbf{EP}och-Wise Dynamic Weighting (EP). To evaluate their effectiveness, we construct four variants: \textbf{EGRA/EBG}, \textbf{EGRA/BDA}, \textbf{EGRA/EN}, and \textbf{EGRA/EP}, which remove the EBG module, the entire BDA module, the EN sub-module, and the EP sub-module, respectively.
We conduct the ablation study on three representative datasets: Sports, Clothing, and MicroLens. The corresponding results are presented in Table~\ref{tab:ae}.
 
\begin{table}
    \caption{Performance comparison of EGRA with its component-related variants.}
    \label{tab:ae}
    \centering
    \footnotesize

    \tabcolsep=0.03cm
    \renewcommand{\arraystretch}{0.6} 
    \begin{tabular}{llccccc}
        \toprule
        Datasets& Metrics & \textbf{EGRA} & \textbf{EGRA/EBG} & \textbf{EGRA/BDA} & \textbf{EGRA/EN} & \textbf{EGRA/EP}\\
        \midrule

        \multirow{4}{*}{Sports}   &R@10 &\textbf{0.0819} &0.0785 &0.0794 &0.0806 &0.0810 \\
                                  &R@20 &\textbf{0.1213} &0.1178 &0.1192 &0.1197 &0.1193 \\
                                  &N@10   &\textbf{0.0449} &0.0431 &0.0433 &0.0445 &0.0441 \\
                                  &N@20   &\textbf{0.0551} &0.0532 &0.0536 &0.0544 &0.0539 \\
        \midrule

        \multirow{4}{*}{Clothing}     &R@10 &\textbf{0.0714} &0.0682 &0.0684 &0.0704 &0.0693 \\
                                  &R@20 &\textbf{0.1043} &0.0987 &0.1005 &0.1031 &0.1025 \\
                                  &N@10   &\textbf{0.0390} &0.0373 &0.0373 &0.0385 &0.0379 \\
                                  &N@20   &\textbf{0.0473} &0.0450 &0.0454 &0.0469 &0.0463 \\
        \midrule

        \multirow{4}{*}{MicroLens} &R@10 &\textbf{0.0890} &0.0863 &0.0872 &0.0881 &0.0886 \\
                                  &R@20 &\textbf{0.1302} &0.1275 &0.1284 &0.1297 &0.1298 \\
                                  &N@10   &\textbf{0.0469} &0.0456 &0.0458 &0.0463 &0.0465 \\
                                  &N@20   &\textbf{0.0575} &0.0562 &0.0564 &0.0571 &0.0571 \\
        \bottomrule
    \end{tabular}
    \vspace{-20pt} 
\end{table}



From Table~\ref{tab:ae}, we first find that, compared with EGRA/EBG and EGRA/BDA, EGRA consistently achieves better performance, indicating the effectiveness of each component in EGRA. Usually, the EBG component emerges as the most significant contributor, followed by BDA. Specifically, compared to EGRA/EBG, EGRA achieves improvements on R@10 by 4.33\%, 4.69\%, and 3.13\% on the Sports, Clothing, and Microlens datasets, respectively. In terms of average gains across the three datasets, EGRA outperforms on R@20, N@10, and N@20 by 3.56\%, 3.86\%, and 3.66\%, respectively. All these substantial improvements demonstrate the effectiveness of the proposed EBG component in enhancing recommendation performance. Compared to EGRA/BDA, EGRA achieves improvements on R@10 by 1.59\%, 4.39\%, and 2.06\% across the three datasets, respectively. For R@20, N@10, and N@20, the average gains reach 2.31\%, 3.70\%, and 2.98\%, respectively. These consistent improvements highlight BDA's effectiveness. 


Second, by comparing the performance of EGRA with EGRA/EN and EGRA/EP, we fine-grainedly observe that both the two sub-components of BDA contribute to the performance gain. Specifically, compared to EGRA/EN, EGRA shows performance improvements on the Sports dataset by 1.61\%, 1.34\%, 0.90\%, and 1.29\% across the four metrics. For the Clothing dataset, the gains are 1.42\%, 1.16\%, 1.30\%, and 0.85\%. On the MicroLens dataset, the improvements reach 1.02\%, 0.39\%, 1.29\%, and 0.70\%. These performance gains demonstrate the effectiveness of the EN sub-component within BDA. Compared to EGRA/EP, EGRA achieves performance gains on the Sports dataset of 1.11\%, 1.68\%, 1.81\%, and 2.23\% across the four metrics. For the Clothing dataset, the improvements are 3.03\%, 1.76\%, 2.90\%, and 2.16\%. On the MicroLens, the corresponding gains are 0.45\%, 0.31\%, 0.86\%, and 0.70\%, respectively. These performance gains demonstrate the effectiveness of the EP sub-component within BDA.

\subsection{Comparison of Different BGE Methods (\textbf{RQ4})}
To further evaluate the effectiveness and generalizability of our Behavior Graph Enhancement strategy, we compare it against the enhancement technique introduced in the recent method GUME, which constructs modality-specific item-item semantic graphs, intersects them to form a unified semantic structure, and integrates the result into the behavior graph.

Specifically, we integrate the enhancement strategies of \textbf{E}GRA and \textbf{G}UME into three representative recommendation models (LightGCN, LATTICE, and MGCN), yielding the variants ``Model+E'' and ``Model+G'', respectively. The evaluation results on four benchmark datasets are presented in Table \ref{tab:cbge1}. 

\begin{table}
    \caption{Performance comparison between the Behavior Graph Enhancement method proposed in EGRA and that of GUME on metric R@20.}
    \label{tab:cbge1}
    \centering

    \renewcommand{\arraystretch}{0.8} 
    \begin{tabular}{llcccc}
        \toprule
        Model& Metrics & \textbf{Baby} & \textbf{Sports} & \textbf{Clothing} & \textbf{Microlens} \\
        \midrule
        LightGCN    &R@20 &0.0754 &0.0864 &0.0544 &0.1075\\
        LightGCN+G  &R@20 &0.0783 &0.0900 &0.0612 &0.1093\\
        LightGCN+E  &R@20 &\textbf{0.0891} &\textbf{0.0997} &\textbf{0.0722} &\textbf{0.1163}\\
        \midrule

        LATTICE    &R@20 &0.0844 &0.0959 &0.0688 &0.1089\\
        LATTICE+G  &R@20 &0.0862 &0.0977 &0.0769 &0.1120\\
        LATTICE+E  &R@20 &\textbf{0.0891} &\textbf{0.1031} &\textbf{0.0834} &\textbf{0.1154}\\
        \midrule

        MGCN    &R@20 &0.0950 &0.1107 &0.0963 &0.1134\\
        MGCN+G  &R@20 &0.0957 &0.1116 &0.0978 &0.1132\\
        MGCN+E  &R@20 &\textbf{0.0991} &\textbf{0.1131} &\textbf{0.0995} &\textbf{0.1157}\\
        \bottomrule
    \end{tabular}
    \vspace{-16pt} 
\end{table}

From the table, it can be observed that the proposed behavior graph enhancement method consistently outperforms the enhancement strategy used in GUME and demonstrates even greater improvements over the variant without any graph enhancement, across all evaluated models and datasets. Specifically, when applied to the base model LightGCN, our strategy achieves significantly larger improvements in R@20 compared to GUME’s strategy—showing gains of 14.32\%, 11.22\%, 20.22\%, and 6.51\% on the Baby, Sports, Clothing, and Microlens datasets, respectively. For the base model LATTICE, our strategy also demonstrates superior performance, achieving improvements of 3.43\%, 5.63\%, 9.45\% and 3.12\% on the same datasets. In the case of MGCN, GUME’s strategy results in only marginal or negligible improvements, while our method consistently outperforms the base model with gains of 4.32\%, 2.17\%, 3.32\%, and 2.03\% across the four datasets. These consistent improvements demonstrate the generalization ability and effectiveness of our proposed BGE strategy.

The superior performance of our BGE method over GUME's primarily stems from two key differences. First, while GUME directly computes item similarity based on raw modality features, our method utilizes item embeddings derived from a pretrained MMR model. These embeddings are optimized to capture both user preference patterns and semantic signals, resulting in more informative and task-relevant representations. Second, the similarity computation in our approach is conducted in the learned embedding space, which is more robust than raw feature space. This reduces the influence of noise and irrelevant modalities, enabling the construction of more semantically meaningful item-item connections. Together, these advantages lead to better graph augmentation and downstream recommendation performance.
\vspace{-20pt}
\subsection{Evaluation with Different Pre-trained Models (\textbf{RQ5})} \label{spm}

In EGRA, MGCN is used as the pre-training backbone. To investigate the potential performance gains from adopting more advanced pre-training models, we perform supplementary experiments by replacing MGCN with GUME, a recent model with state-of-the-art recommendation performance.  The results are reported in Table \ref{pcpb}.

\begin{table}[h]
\caption{Performance comparison under different \textbf{p}re-\textbf{t}raining \textbf{m}odel (PTM). The \textbf{bold} figures represent the best results among all methods.}\label{pcpb}
\footnotesize
\setlength{\tabcolsep}{2mm}{
\begin{tabular}{llccccc}
\toprule
\multirow{2}{*}{PTM} & \multirow{2}{*}{Metric} & \multicolumn{5}{c}{Dataset} \\
\cmidrule(lr){3-7}
& & Baby & Sports & Clothing & Microlens & Elec \\
\midrule

MGCN & \multirow{2}{*}{R@10} &0.0703  &0.0819  &0.0714  &0.0890  &0.0498  \\
GUME &                       &\textbf{0.0707}  &\textbf{0.0822}  &\textbf{0.0721}  &\textbf{0.0898}  &\textbf{0.0503}  \\
\midrule
MGCN & \multirow{2}{*}{R@20} &0.1058  &0.1213  &0.1043  &0.1302  &0.0732  \\
GUME &                       &\textbf{0.1064}  &\textbf{0.1220}  &\textbf{0.1046}  &\textbf{0.1312}  &\textbf{0.0737}  \\
\midrule
MGCN & \multirow{2}{*}{N@10} &0.0379  &0.0449  &0.0390  &0.0469  &0.0282  \\
GUME &                       &\textbf{0.0381}  &\textbf{0.0451}  &\textbf{0.0392}  &\textbf{0.0472}  &\textbf{0.0284}  \\
\midrule
MGCN & \multirow{2}{*}{N@20} &\textbf{0.0470}  &0.0551  &0.0473  &0.0575  &0.0343  \\
GUME &                       &\textbf{0.0470}  &\textbf{0.0553}  &\textbf{0.0474}  &\textbf{0.0579}  &\textbf{0.0344}  \\

\bottomrule
\end{tabular}}
\end{table}

From Table~\ref{pcpb}, we observe that although GUME substantially outperforms MGCN as a standalone model, replacing MGCN with GUME yields only marginal gains for EGRA.

A possible explanation for this observation lies in the role of the pretrained model in EGRA. Specifically, the representations from the pretrained model are not used to directly optimize recommendation performance, but rather to induce item–item relations that enrich the original user–item interaction graph. Improving the standalone accuracy of the pretrained model may refine the induced item–item relations. However, once the recommendation accuracy reaches a certain level, such refinements do not necessarily introduce substantially new structural information for downstream graph-based representation learning, and therefore may not lead to noticeable performance gains in EGRA.

When training resources are constrained and extreme performance improvements are not required, MGCN is a more feasible pre-training backbone than the more complex GUME.

\vspace{-5pt} 
\subsection{Comparison of Training Efficiency (\textbf{RQ6})}

We also compare the training efficiency of EGRA, MGCN, and the best-performing baseline, GUME, in Table~\ref{tab:ptt}.

\begin{table}[H]
    \caption{Per-epoch training time comparison (s).}
    \label{tab:ptt}
    \centering
    \def\arraystretch{1.23}	
    \tabcolsep=0.27cm
    \renewcommand{\arraystretch}{1.25} 
    \begin{tabular}{lccccc}
        \toprule
        \multirow{2}{*}{Models} & \multicolumn{5}{c}{Datasets}\\
        \cline{2-6}
        & Baby & Sports & Clothing & Microlens & Elec\\
        \midrule
        EGRA &1.47 &3.34 &3.21 &9.28 &56.92 \\
        MGCN &1.19 &2.86 &2.83 &7.71 &48.07 \\
        GUME &2.04 &5.01 &4.55 &15.73 &96.62 \\
        \bottomrule
    \end{tabular}
    \vspace{-8pt} 
\end{table}

First, as shown in Table~\ref{tab:ptt}, EGRA achieves lower training cost than GUME while incurring only a modest increase in training time compared with MGCN. Specifically, our model requires only 72.05\%, 66.67\%, 70.55\%, 59.00\%, and 58.91\% of GUME’s per-epoch training time across the five datasets, respectively. The lower training cost of EGRA compared with GUME stems from its more efficient architecture. While EGRA introduces additional alignment computation, it performs graph convolution only once, whereas GUME performs graph convolution three times and additionally constructs contrastive views through noise injection. As a result, GUME incurs significantly higher computational overhead. Compared with MGCN, the slightly increased training cost of EGRA is mainly due to the enhanced behavior graph and the representation alignment mechanism.   

In addition, because EGRA relies on pretrained representations from MGCN, we further consider the cost of the MGCN pre-training stage when evaluating the overall training overhead. Given the similar convergence behavior across methods, the overall overhead is estimated by summing the training costs of MGCN and EGRA. By combining the results of EGRA and MGCN in Table~\ref{tab:ptt}, the overall training time is observed to be moderately higher than that of GUME, with increases of 30.39\%, 23.75\%, 32.75\%, 8.01\%, and 8.66\% on Baby, Sports, Clothing, Microlens, and Elec, respectively. Overall, the relative increase in training time introduced by EGRA decreases as the dataset size grows. Notably, the additional training time on the two larger datasets (Microlens and Elec) remains below 9\%, indicating that the practical overhead introduced by the pre-training strategy becomes increasingly marginal in large-scale recommendation scenarios. Therefore, despite adopting a pre-training strategy, EGRA still maintains efficient training behavior and good scalability. Jointly considering EGRA’s significant performance advantage and its slight increase in overall time cost, especially on large datasets, the method exhibits promising practical applicability.

\subsection{Model Parameter Analysis (\textbf{RQ7})}

We conduct a parameter analysis on two key hyperparameters of EGRA: the number of neighboring items $H$ used for constructing the item–item graph, and the initial epoch-wise alignment weight $\lambda_{min}$.

\begin{figure}[h]
    \centering
\includegraphics[width=0.48\textwidth]{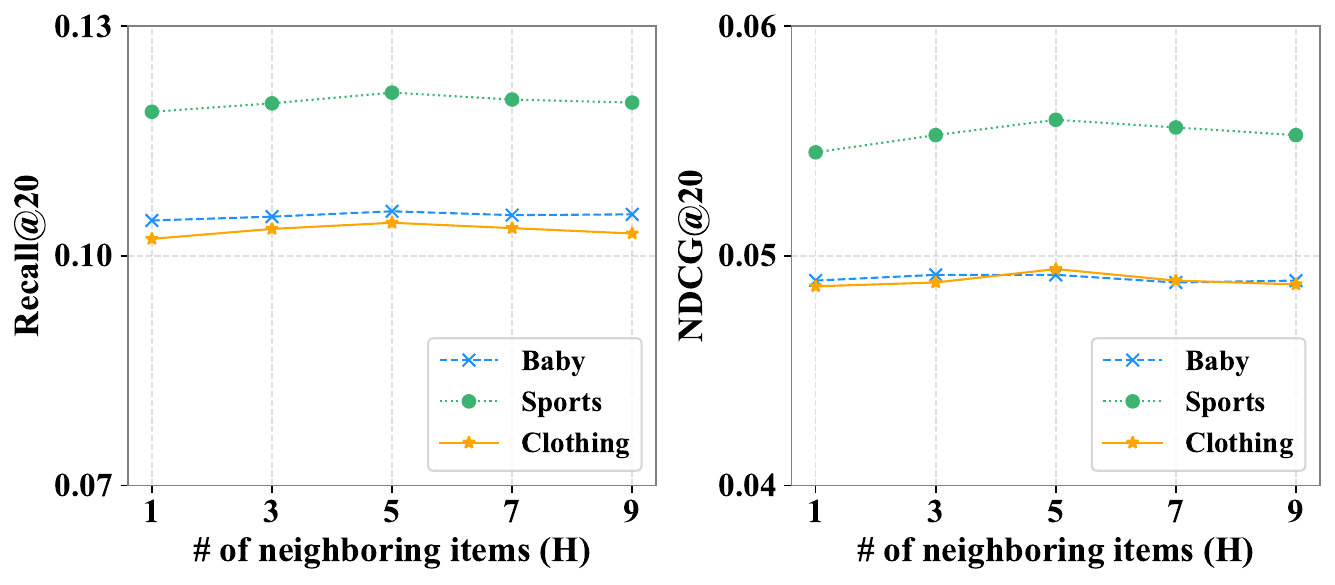} 
    \caption{Performance of \textbf{EGRA} with respect to varying $H$.}  
    \label{fig:para_h}  
\end{figure}

The number of neighboring items $H$ is selected from $\{1, 3, 5, 7, 9\}$, and the results are shown in Figure~\ref{fig:para_h}. From it, we find that the performance of EGRA on both R@20 and N@20 generally follows a rising-then-falling pattern as the number of neighboring items $H$ increases from 1 to 9. This trend is especially evident on the Sports and Clothing datasets. For instance, on Sports, R@20 improves steadily from 0.1188 to a peak of 0.1213 at $H = 5$, before slightly declining to 0.1200 at $H = 9$. A similar pattern is observed for N@20. The Clothing dataset exhibits the same trend, with optimal values also occurring when $H = 5$. In contrast, the Baby dataset is less sensitive to variations in $H$, with only marginal performance differences across different settings, although the best performance is still achieved at $H=5$. This observation is consistent with the results in Table~\ref{tab:me}, where Baby exhibits relatively smaller gains from behavior graph enhancement than the other datasets.



\begin{table}[H]
    \caption{Performance variation of EGRA as the different initial epoch-wise alignment weight $\lambda_{min}$.}
    \label{tab:vea}
    \centering
    \tabcolsep=0.30cm
    \renewcommand{\arraystretch}{1.0} 
    \begin{tabular}{llccc}
        \toprule
        \multirow{2}{*}{Dataset} & \multirow{2}{*}{Metrics} & \multicolumn{3}{c}{$\lambda_{min}$} \\
        \cline{3-5}
        
        &&0.0 & 0.005 & 0.01 \\
        \midrule
        \multirow{2}{*}{Sports}      &R@20 &0.1128 &0.1203 &0.1213\\
                                     &N@20 &0.0508 &0.0547 &0.0551\\
        \midrule
        \multirow{2}{*}{Clothing}    &R@20 &0.1024 &0.1028 &0.1043\\
                                     &N@20 &0.0463 &0.0466 &0.0473\\
        \midrule

        \multirow{2}{*}{Microlens}    &R@20 &0.1258 &0.1302 &0.1299\\
                                      &N@20 &0.0558 &0.0575 &0.0574\\
        \bottomrule
    \end{tabular} 
\end{table}

The Table \ref{tab:vea} shows how the varying initial epoch-wise alignment weight $\lambda_{min}$ influence the performance of EGRA. From it, we can find that selecting an appropriate value for $\lambda_{min}$ plays a crucial role in achieving strong performance. All datasets show that setting $\lambda_{min} = 0$, i.e., starting the epoch-wise alignment weight from zero, generally leads to suboptimal performance. For example, on the Sports dataset, this setting results in a performance drop of 7.54\% on R@20 and 8.46\% on N@20 compared to the best configuration. On the Clothing dataset, the drops are 1.86\% and 2.16\%, while on MicroLens, the performance decreases by 3.50\% and 3.05\% on R@20 and N@20, respectively. However, a higher $\lambda_{min}$ does not necessarily lead to better performance. In fact, unlike Sports and Clothing, the MicroLens dataset achieves its best results when $\lambda_{min}$ is set to 0.005 rather than the higher value of 0.01. Hence, selecting an appropriate $\lambda_{min}$ for each dataset is essential for achieving optimal performance.

\section{Conclusion and Discussion}
In this work, we proposed EGRA, a novel multimodal recommendation framework that simultaneously enhances the behavior graph and improves modality–behavior alignment. EGRA introduces a behavior graph enhancement strategy that augments the interaction graph with an item–item graph constructed from pretrained representations, capturing both collaborative and modality-aware similarities while mitigating modality noise. Furthermore, it incorporates a bi-level dynamic alignment weighting mechanism that adaptively adjusts alignment strength across entities and training epochs, enabling more stable and personalized representation alignment. Extensive experiments on five datasets show the effectiveness and generality of EGRA across diverse recommendation scenarios.

While EGRA achieves strong performance, it depends on a separately pre-trained multimodal model to build the item-item semantic graph, incurring extra computation. In future work, we aim to develop a unified and efficient strategy to dynamically construct the semantic graph during training—e.g., by pre-training EGRA for a few epochs, then extracting the item-item graph from intermediate embeddings for joint optimization. This approach could reduce complexity while preserving the benefits of semantic enhancement.


\newpage
\balance
\bibliographystyle{IEEEtran}
\bibliography{references}

@inproceedings{gume,
  title={GUME: Graphs and User Modalities Enhancement for Long-Tail Multimodal Recommendation},
  author={Lin, Guojiao and Zhen, Meng and Wang, Dongjie and Long, Qingqing and Zhou, Yuanchun and Xiao, Meng},
  booktitle={Proceedings of the 33rd ACM International Conference on Information and Knowledge Management},
  pages={1400--1409},
  year={2024}
}

@inproceedings{lin2024gume,
  title={GUME: Graphs and User Modalities Enhancement for Long-Tail Multimodal Recommendation},
  author={Lin, Guojiao and Zhen, Meng and Wang, Dongjie and Long, Qingqing and Zhou, Yuanchun and Xiao, Meng},
  booktitle={Proceedings of the 33rd ACM International Conference on Information and Knowledge Management},
  pages={1400--1409},
  year={2024}
}

@inproceedings{fan2023graph,
  title={Graph collaborative signals denoising and augmentation for recommendation},
  author={Fan, Ziwei and Xu, Ke and Dong, Zhang and Peng, Hao and Zhang, Jiawei and Yu, Philip S},
  booktitle={Proceedings of the 46th international ACM SIGIR conference on research and development in information retrieval},
  pages={2037--2041},
  year={2023}
}

@inproceedings{bu2010music,
  title={Music recommendation by unified hypergraph: combining social media information and music content},
  author={Bu, Jiajun and Tan, Shulong and Chen, Chun and Wang, Can and Wu, Hao and Zhang, Lijun and He, Xiaofei},
  booktitle={Proceedings of the 18th ACM international conference on Multimedia},
  pages={391--400},
  year={2010}
}

@article{liu2022multimodal,
  title={Multimodal hierarchical graph collaborative filtering for multimedia-based recommendation},
  author={Liu, Kang and Xue, Feng and Li, Shuaiyang and Sang, Sheng and Hong, Richang},
  journal={IEEE Transactions on Computational Social Systems},
  volume={11},
  number={1},
  pages={216--227},
  year={2022},
  publisher={IEEE}
}

@article{slmrec,
  title={Self-supervised learning for multimedia recommendation},
  author={Tao, Zhulin and Liu, Xiaohao and Xia, Yewei and Wang, Xiang and Yang, Lifang and Huang, Xianglin and Chua, Tat-Seng},
  journal={IEEE Transactions on Multimedia},
  volume={25},
  pages={5107--5116},
  year={2022},
  publisher={IEEE}
}

@inproceedings{zhou2023mmrec,
  title={Mmrec: Simplifying multimodal recommendation},
  author={Zhou, Xin},
  booktitle={Proceedings of the 5th ACM International Conference on Multimedia in Asia Workshops},
  pages={1--2},
  year={2023}
}

@inproceedings{xv2024improving,
  title={Improving Multi-modal Recommender Systems by Denoising and Aligning Multi-modal Content and User Feedback},
  author={Xv, Guipeng and Li, Xinyu and Xie, Ruobing and Lin, Chen and Liu, Chong and Xia, Feng and Kang, Zhanhui and Lin, Leyu},
  booktitle={Proceedings of the 30th ACM SIGKDD Conference on Knowledge Discovery and Data Mining},
  pages={3645--3656},
  year={2024}
}

@inproceedings{vbpr,
author = {He, Ruining and McAuley, Julian},
title = {VBPR: visual Bayesian Personalized Ranking from implicit feedback},
year = {2016},
publisher = {AAAI Press},
booktitle = {Proceedings of the Thirtieth AAAI Conference on Artificial Intelligence},
pages = {144–150},
numpages = {7},
series = {AAAI'16}
}

@article{dvbpr,
  title={Visually-Aware Fashion Recommendation and Design with Generative Image Models},
  author={Wang-Cheng Kang and Chen Fang and Zhaowen Wang and Julian McAuley},
  journal={2017 IEEE International Conference on Data Mining (ICDM)},
  year={2017},
  pages={207-216}
}

@inproceedings{vecf,
author = {Chen, Xu and Chen, Hanxiong and Xu, Hongteng and Zhang, Yongfeng and Cao, Yixin and Qin, Zheng and Zha, Hongyuan},
title = {Personalized Fashion Recommendation with Visual Explanations based on Multimodal Attention Network: Towards Visually Explainable Recommendation},
year = {2019},
isbn = {9781450361729},
publisher = {Association for Computing Machinery},
booktitle = {Proceedings of the 42nd International ACM SIGIR Conference on Research and Development in Information Retrieval},
pages = {765–774},
numpages = {10},
series = {SIGIR'19}
}

@inproceedings{mmgcn,
author = {Wei, Yinwei and Wang, Xiang and Nie, Liqiang and He, Xiangnan and Hong, Richang and Chua, Tat-Seng},
title = {MMGCN: Multi-modal Graph Convolution Network for Personalized Recommendation of Micro-video},
year = {2019},
isbn = {9781450368896},
publisher = {Association for Computing Machinery},
booktitle = {Proceedings of the 27th ACM International Conference on Multimedia},
pages = {1437–1445},
numpages = {9},
series = {MM '19}
}

@inproceedings{grcn,
author = {Wei, Yinwei and Wang, Xiang and Nie, Liqiang and He, Xiangnan and Chua, Tat-Seng},
year = {2020},
month = {10},
pages = {3541-3549},
title = {Graph-Refined Convolutional Network for Multimedia Recommendation with Implicit Feedback},
doi = {10.1145/3394171.3413556}
}

@article{dualgcn,
author = {Wang, Qifan and Wei, Yinwei and Yin, Jianhua and Wu, Jianlong and Song, Xuemeng and Nie, Liqiang},
title = {DualGNN: Dual Graph Neural Network for Multimedia Recommendation},
year = {2023},
issue_date = {2023},
publisher = {IEEE Press},
volume = {25},
issn = {1520-9210},
journal = {Trans. Multi.},
month = jan,
pages = {1074–1084},
numpages = {11}
}

@inproceedings{mgcn,
author = {Yu, Penghang and Tan, Zhiyi and Lu, Guanming and Bao, Bing-Kun},
title = {Multi-View Graph Convolutional Network for Multimedia Recommendation},
year = {2023},
isbn = {9798400701085},
publisher = {Association for Computing Machinery},
address = {New York, NY, USA},
booktitle = {Proceedings of the 31st ACM International Conference on Multimedia},
pages = {6576–6585},
numpages = {10},
location = {Ottawa ON, Canada},
series = {MM '23}
}

@inproceedings{lattice,
author = {Zhang, Jinghao and Zhu, Yanqiao and Liu, Qiang and Wu, Shu and Wang, Shuhui and Wang, Liang},
title = {Mining Latent Structures for Multimedia Recommendation},
year = {2021},
isbn = {9781450386517},
publisher = {Association for Computing Machinery},
address = {New York, NY, USA},
booktitle = {Proceedings of the 29th ACM International Conference on Multimedia},
pages = {3872–3880},
numpages = {9},
location = {Virtual Event, China},
series = {MM '21}
}

@article{micro,
author = {Zhang, Jinghao and Zhu, Yanqiao and Liu, Qiang and Zhang, Mengqi and Wu, Shu and Wang, Liang},
title = {Latent Structure Mining With Contrastive Modality Fusion for Multimedia Recommendation},
year = {2023},
issue_date = {Sept. 2023},
publisher = {IEEE Educational Activities Department},
address = {USA},
volume = {35},
number = {9},
issn = {1041-4347},
journal = {IEEE Trans. on Knowl. and Data Eng.},
month = sep,
pages = {9154–9167},
numpages = {14}
}

@inproceedings{freedom,
author = {Zhou, Xin and Shen, Zhiqi},
title = {A Tale of Two Graphs: Freezing and Denoising Graph Structures for Multimodal Recommendation},
year = {2023},
isbn = {9798400701085},
publisher = {Association for Computing Machinery},
booktitle = {Proceedings of the 31st ACM International Conference on Multimedia},
pages = {935–943},
numpages = {9},
series = {MM '23}
}

@inproceedings{bm3,
author = {Zhou, Xin and Zhou, Hongyu and Liu, Yong and Zeng, Zhiwei and Miao, Chunyan and Wang, Pengwei and You, Yuan and Jiang, Feijun},
title = {Bootstrap Latent Representations for Multi-modal Recommendation},
year = {2023},
isbn = {9781450394161},
publisher = {Association for Computing Machinery},
booktitle = {Proceedings of the ACM Web Conference 2023},
pages = {845–854},
numpages = {10},
series = {WWW '23}
}

@article{lgmrec,
author = {Guo, Zhiqiang and Li, Jianjun and Guohui, Li and Wang, Chaoyang and Shi, Si and Ruan, Bin},
year = {2024},
month = {03},
pages = {8454-8462},
title = {LGMRec: Local and Global Graph Learning for Multimodal Recommendation},
volume = {38},
journal = {Proceedings of the AAAI Conference on Artificial Intelligence}
}

@inproceedings{lightgcn,
  title={Lightgcn: Simplifying and powering graph convolution network for recommendation},
  author={He, Xiangnan and Deng, Kuan and Wang, Xiang and Li, Yan and Zhang, Yongdong and Wang, Meng},
  booktitle={Proceedings of the 43rd International ACM SIGIR conference on research and development in Information Retrieval},
  pages={639--648},
  year={2020}
}

@article{chen2009fast,
  title={Fast Approximate kNN Graph Construction for High Dimensional Data via Recursive Lanczos Bisection.},
  author={Chen, Jie and Fang, Haw-ren and Saad, Yousef},
  journal={Journal of Machine Learning Research},
  volume={10},
  number={9},
  year={2009}
}

@article{ni2023content,
  title={A content-driven micro-video recommendation dataset at scale},
  author={Ni, Yongxin and Cheng, Yu and Liu, Xiangyan and Fu, Junchen and Li, Youhua and He, Xiangnan and Zhang, Yongfeng and Yuan, Fajie},
  journal={arXiv preprint arXiv:2309.15379},
  year={2023}
}

@article{xu2021multi,
  title={Multi-modal discrete collaborative filtering for efficient cold-start recommendation},
  author={Xu, Yang and Zhu, Lei and Cheng, Zhiyong and Li, Jingjing and Zhang, Zheng and Zhang, Huaxiang},
  journal={IEEE Transactions on Knowledge and Data Engineering},
  volume={35},
  number={1},
  pages={741--755},
  year={2021},
  publisher={IEEE}
}

@inproceedings{liu2017deepstyle,
  title={Deepstyle: Learning user preferences for visual recommendation},
  author={Liu, Qiang and Wu, Shu and Wang, Liang},
  booktitle={Proceedings of the 40th international acm sigir conference on research and development in information retrieval},
  pages={841--844},
  year={2017}
}

@article{baltruvsaitis2018multimodal,
  title={Multimodal machine learning: A survey and taxonomy},
  author={Baltru{\v{s}}aitis, Tadas and Ahuja, Chaitanya and Morency, Louis-Philippe},
  journal={IEEE transactions on pattern analysis and machine intelligence},
  volume={41},
  number={2},
  pages={423--443},
  year={2018},
  publisher={IEEE}
}

@article{yin2012challenging,
  title={Challenging the Long Tail Recommendation},
  author={Yin, Hongzhi and Cui, Bin and Li, Jing and Yao, Junjie and Chen, Chen},
  journal={Proceedings of the VLDB Endowment},
  volume={5},
  number={9},
  year={2012}
}

@article{volkovs2017dropoutnet,
  title={Dropoutnet: Addressing cold start in recommender systems},
  author={Volkovs, Maksims and Yu, Guangwei and Poutanen, Tomi},
  journal={Advances in neural information processing systems},
  volume={30},
  year={2017}
}

@inproceedings{lee2019melu,
  title={Melu: Meta-learned user preference estimator for cold-start recommendation},
  author={Lee, Hoyeop and Im, Jinbae and Jang, Seongwon and Cho, Hyunsouk and Chung, Sehee},
  booktitle={Proceedings of the 25th ACM SIGKDD international conference on knowledge discovery \& data mining},
  pages={1073--1082},
  year={2019}
}

@inproceedings{liu2020long,
  title={Long-tail session-based recommendation},
  author={Liu, Siyi and Zheng, Yujia},
  booktitle={Proceedings of the 14th ACM conference on recommender systems},
  pages={509--514},
  year={2020}
}

@inproceedings{zhang2021model,
  title={A model of two tales: Dual transfer learning framework for improved long-tail item recommendation},
  author={Zhang, Yin and Cheng, Derek Zhiyuan and Yao, Tiansheng and Yi, Xinyang and Hong, Lichan and Chi, Ed H},
  booktitle={Proceedings of the web conference 2021},
  pages={2220--2231},
  year={2021}
}

@inproceedings{luo2023improving,
  title={Improving long-tail item recommendation with graph augmentation},
  author={Luo, Sichun and Ma, Chen and Xiao, Yuanzhang and Song, Linqi},
  booktitle={Proceedings of the 32nd ACM international conference on information and knowledge management},
  pages={1707--1716},
  year={2023}
}

@inproceedings{wu2024coral,
  title={Coral: Collaborative retrieval-augmented large language models improve long-tail recommendation},
  author={Wu, Junda and Chang, Cheng-Chun and Yu, Tong and He, Zhankui and Wang, Jianing and Hou, Yupeng and McAuley, Julian},
  booktitle={Proceedings of the 30th ACM SIGKDD Conference on Knowledge Discovery and Data Mining},
  pages={3391--3401},
  year={2024}
}

@article{liu2024llm,
  title={Llm-esr: Large language models enhancement for long-tailed sequential recommendation},
  author={Liu, Qidong and Wu, Xian and Wang, Yejing and Zhang, Zijian and Tian, Feng and Zheng, Yefeng and Zhao, Xiangyu},
  journal={Advances in Neural Information Processing Systems},
  volume={37},
  pages={26701--26727},
  year={2024}
}

@article{luo2024integrating,
  title={Integrating large language models into recommendation via mutual augmentation and adaptive aggregation},
  author={Luo, Sichun and Yao, Yuxuan and He, Bowei and Huang, Yinya and Zhou, Aojun and Zhang, Xinyi and Xiao, Yuanzhang and Zhan, Mingjie and Song, Linqi},
  journal={arXiv preprint arXiv:2401.13870},
  year={2024}
}

@inproceedings{liu2023diffusion,
  title={Diffusion augmentation for sequential recommendation},
  author={Liu, Qidong and Yan, Fan and Zhao, Xiangyu and Du, Zhaocheng and Guo, Huifeng and Tang, Ruiming and Tian, Feng},
  booktitle={Proceedings of the 32nd ACM International conference on information and knowledge management},
  pages={1576--1586},
  year={2023}
}

@inproceedings{liu2023linrec,
  title={Linrec: Linear attention mechanism for long-term sequential recommender systems},
  author={Liu, Langming and Cai, Liu and Zhang, Chi and Zhao, Xiangyu and Gao, Jingtong and Wang, Wanyu and Lv, Yifu and Fan, Wenqi and Wang, Yiqi and He, Ming and others},
  booktitle={Proceedings of the 46th International ACM SIGIR Conference on Research and Development in Information Retrieval},
  pages={289--299},
  year={2023}
}

@article{zhou2025learning,
  title={Learning Item Representations Directly from Multimodal Features for Effective Recommendation},
  author={Zhou, Xin and Zhang, Xiaoxiong and Niyato, Dusit and Shen, Zhiqi},
  journal={arXiv preprint arXiv:2505.04960},
  year={2025}
}

@article{zhang2025semantic,
  title={Semantic Item Graph Enhancement for Multimodal Recommendation},
  author={Zhang, Xiaoxiong and Zhou, Xin and Zeng, Zhiwei and Niyato, Dusit and Shen, Zhiqi},
  journal={arXiv preprint arXiv:2508.06154},
  year={2025}
}

@article{fedc,
  title={Communication-Efficient Federated Knowledge Graph Embedding with Entity-Wise Top-K Sparsification},
  author={Zhang, Xiaoxiong and Zeng, Zhiwei and Zhou, Xin and Niyato, Dusit and Shen, Zhiqi},
  journal={Knowledge-Based Systems},
  pages={114147},
  year={2025},
  publisher={Elsevier}
}

\vfill

\end{document}